\documentclass[canadian,english]{spie}
\usepackage[T1]{fontenc}
\usepackage[latin9]{inputenc}
\usepackage{float}
\usepackage{units}
\usepackage{graphicx}
\usepackage{setspace}

\makeatletter
\newcommand{\lyxaddress}[1]{
\par {\raggedright #1
\vspace{1.4em}
\noindent\par}
}

\usepackage{graphicx}

\@ifundefined{showcaptionsetup}{}{%
 \PassOptionsToPackage{caption=false}{subfig}}
\usepackage{subfig}
\makeatother

\usepackage{babel}
\begin{document}

\title{{\large{}Can A Universal Quantum Cloner Be Used to Design an Experimentally
Feasible Near-Deterministic CNOT Gate ?}}

\author{{\small{}Amor Gueddana$^{1,\,2}$ , Peyman Gholami$^{1}$ and Vasudevan
Lakshminarayanan$^{1,\,3}$}}

\maketitle

\lyxaddress{1. Theoretical \& Experimental Epistemology Lab, TEEL, School of
Optometry and Vision Science, University of Waterloo 200 University
Avenue West, Waterloo, Ontario N2l 3G1, Canada.}

\lyxaddress{2. Green \& Smart Communication Systems Lab, Gres'Com, Engineering
School of Communication of Tunis, Sup'Com, University of Carthage,
Ghazela Technopark, 2083, Ariana, Tunisia.}

\lyxaddress{3. Department of Physics, Department of Electrical and Computer Engineering
and Department of Systems Design Engineering , University of Waterloo
200 University Avenue West, Waterloo, Ontario N2l 3G1, Canada.}
\begin{abstract}
We propose a non-deterministic CNOT gate based on a quantum cloner,
a quantum switch based on all optical routing of single photon by
single photon, a quantum-dot spin in a double-sided optical microcavity
with two photonic qubits, delay lines and other linear optical photonic
devices. Our CNOT provides a fidelity of 78\% with directly useful
outputs for a quantum computing circuit and requires no ancillary
qubits or electron spin measurements.
\end{abstract}

\section{Introduction}

\selectlanguage{canadian}%
Physical implementation of the photonic quantum computer and secure
optical quantum communication systems are based on photonic quantum
gates \cite{Kok_2007}. The gate which is universal for building all
quantum circuits is the Controlled-NOT gate (CNOT) \cite{Shende_2006,Maslov_2008}.
Experimentally realizable photonic CNOT gates are those based on linear
optical devices \cite{Pittman_2001,Ralph_2002,OBrien_2003,Pittman_2003,Gasparoni_2004,Pittman_2004,Bao_2007,Clark_2009,Gueddana_2013a,Gueddana_2015}.
These gates have success probabilities less than 1/4. Further improvement
in the success probability of this CNOT model is not possible because
the best success probability for the CNOT functioning is 3/4 when
using linear optical devices \cite{Knill_2003}. This is a major hurdle
for realization of all complex versions of the quantum computing circuits,
since the success probability of such circuits may be very low due
to multiple combinations of many non-deterministic CNOTs . To overcome
this inefficiency, other techniques such as superconducting qubits
\cite{Plantenberg_2007} have to be employed. Other work using non-linearities
have been proposed in order to achieve non-deterministic CNOTs with
high success probability \cite{Luo_2016,Li_2013}. Other designs of
CNOTs are based on spin of electron trapped in Quantum Dot (QD) and
confined in a double sided optical micro-cavity \cite{Wei_2013,Wei_2014,Wang_2013,Luo_2014,Bonato_2010}.
Although the fidelity values of these CNOTs seems to be near unity,
in practice all these designs have major drawbacks. This is because
of physical constraints that make them less effective in serial or
parallel combinations. For the special CNOT model of Wang \textit{et
al} \cite{Wang_2013}, we presented comments on the parametric values
taken for the simulation and showed that the proposed CNOT gate is
valid only in the strong coupling regime \cite{Gueddana_2018d}. In
this paper, we refer to this same CNOT and show that it provides a
fidelity of only 47\% in a realistic implementation (while it is 93.7\%
in the theoretical case), then we propose an optimized design based
on a quantum cloner. 

Several theoretical and experimental work has addressed quantum cloning
machines providing optimal polarization cloning of single photons
when using either Parametric Down Conversion (PDC) or photon bunching
on Beam Splitter (BS). Fasel \textit{et al \cite{Fasel_2002} }proposed
close-to optimal quantum cloning of the polarization state of light
using standard Erbium dopped fiber for amplification and provided
a fidelity $F_{cloner}$ equal to 0.82. Linares \textit{et al} \cite{Lamas_Linares_2002}
proposed a cloning technique, based on stimulated emission in PDC
using non linear $\beta$-Barium borate (BBo) crystal, and obtained
experimental results with $F_{cloner}=0.81$. Martini \textit{et al}
\cite{Martini_2004} used for cloning a BBo crystal slab cut for type
II phase matching for implementing optimal cloning and NOT gate. Bartuskova
\textit{et al} \cite{Bartifmmodemathringuelserufiifmmodecheckselsevsfikova_2007}
addressed in their work a phase covariant $1\rightarrow2$ qubit cloner
based on two single mode optical fiber, non linear crystal, attenuator
and phase modulator, providing a fidelity of $F_{cloner}=0.854$,
which slightly surpasses the theoretical optimal value of the Universal
Cloner (UC), denoted $F_{UC}=5/6$. The question is: can these cloners
be used for designing photonic CNOT gates? The main objective of this
paper is to answer this question. Therefore, this paper is composed
of five sections: Section 2 describes the several photonic components
used for the CNOT design in the imperfect case. Section 3 presents
the concept and modelling of the photonic CNOT using a quantum cloner
and two quantum switches. In section 4, we present the simulation
results and the corresponding experimental realization challenges.
Finally, a conclusion is given in section 5.

\selectlanguage{english}%

\section{Imperfect photonic devices}

\begin{figure}[H]
\begin{centering}
\subfloat[\label{fig:1a} HWP]{\begin{centering}
\includegraphics[width=0.15\columnwidth]{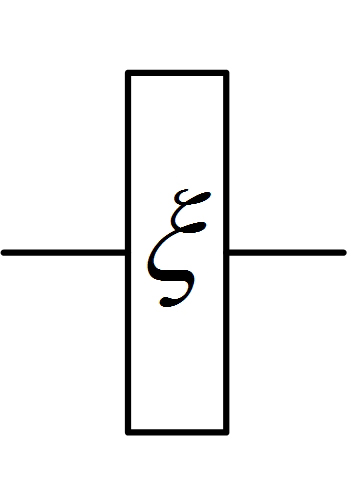}
\par\end{centering}

}\hspace{1cm}\subfloat[\label{fig:1b} CPBS]{\begin{centering}
\includegraphics[width=0.2\columnwidth]{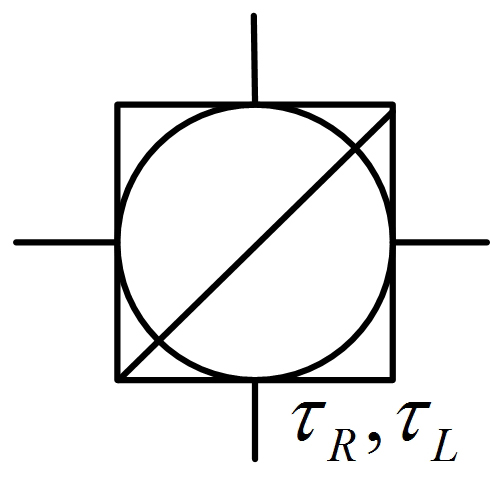}
\par\end{centering}

}\hspace{1cm}\subfloat[\label{fig:1c} SW]{\begin{centering}
\includegraphics[width=0.2\columnwidth]{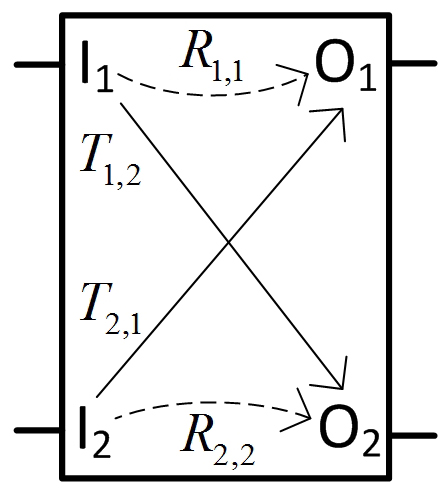}
\par\end{centering}

}
\par\end{centering}

\caption{\label{fig:1}Photonic devices used for the design.}
\end{figure}

\selectlanguage{canadian}%
Let us first consider the basic photonic components in the imperfect
case as illustrated by figure \ref{fig:1}.

\selectlanguage{english}%
Figure \ref{fig:1a} illustrates a Half Wave Plate (HWP) with arbitrary
error{\small{} $\xi$}. For inputs right-circularly polarized single
photon denoted{\small{}$\left|R\right\rangle $ }and left-circularly
polarized single photon denoted{\small{}$\left|L\right\rangle $,}
the HWP behaves as follows:

{\small{}
\begin{equation}
\begin{array}{c}
\left|R\right\rangle \rightarrow\sqrt{\frac{1-\xi}{2}}\left|R\right\rangle +\sqrt{\frac{1+\xi}{2}}\left|L\right\rangle ;\,\,\,\left|L\right\rangle \rightarrow\sqrt{\frac{1-\xi}{2}}\left|R\right\rangle -\sqrt{\frac{1+\xi}{2}}\left|L\right\rangle \end{array}\label{eq:1}
\end{equation}
}{\small \par}

An ideal Circular Polarizing Beam Splitter (CPBS) transmits{\small{}
$\left|R\right\rangle $} and totally reflects{\scriptsize{} }{\small{}$\left|L\right\rangle $.}
When considering a CPBS with arbitrary errors{\small{} $\tau_{R}$}
and {\small{}$\tau_{L}$ }on {\small{}$\left|R\right\rangle $} and
{\small{}$\left|L\right\rangle $,} and for an arbitrary incident
input state{\small{} $\alpha\left|R\right\rangle +\beta\left|L\right\rangle $}
(figure \ref{fig:1b}), the CPBS transmits the state{\small{} $\sqrt{\alpha-\tau_{R}}\left|R\right\rangle +\sqrt{\tau_{L}}\left|L\right\rangle $
}and reflects the state {\small{}$\sqrt{\tau_{R}}\left|R\right\rangle +\sqrt{\beta-\tau_{L}}\left|L\right\rangle $.}{\small \par}

A quantum switch (SW) has two inputs $I_{1}$ and $I_{2}$, and two
outputs $O_{1}$ and $O_{2}$. The transmittance ( solid arrows in
figure\ref{fig:1c}) and reflectance ( dashed arrows in figure\ref{fig:1c})
coefficients from $I_{1}$ and $I_{2}$ to $O_{1}$ and $O_{2}$,
are denoted by $T_{1,2}$, $T_{2,1}$, $R_{1,1}$ and $R_{2,2}$.
Several works addressed physical implementation of the quantum SW
\cite{OShea_2013,Smart_2014,Sun_2016,Volz_2012}. The switching process
considered in this work is based on all optical routing of single
photon by single photon, without any additional control field \cite{Shomroni_2014}.
This switch is based on three-level atomic $\Lambda$-configuration,
with two different transitions denoted $\sigma^{+}$ and $\sigma^{-}$,
and representing transitions coupled only to the right or to the left
photonic mode propagation, respectively. Each of the inputs $I_{1}$
or $I_{2}$, may be either transmitted or reflected into $O_{1}$
or $O_{2}$, depending on the atom's state, being either on the $m_{F}=+1$
state or $m_{F}=-1$. When the atom is in this latter state, incoming
photon on $I_{1}$ in the state $\sigma^{+}$ are reflected to $O_{1}$
and their state becomes $\sigma^{-}$, which toggles the atom's state
to $m_{F}=+1$. When the atom's state is $m_{F}=+1$, it doesn't interact
with $\sigma^{+}$ photons from $I_{1}$ and they are totally transmitted
to $O_{2}$. The whole process is symmetric for $I_{2}$. 

Other photonic devices used in our optimized model are the Beam Splitter
(BS), the Quarter Wave Plate (QWP) and Delay Line (DL). In this work,
we consider these devices only in the ideal case.

\section{Optimized CNOT gate model based on a quantum cloner }

With the previously mentioned components , we propose a CNOT architecture
using the quantum universal cloner as illustrated by figure \ref{fig:2}.

The central part of the CNOT shaded grey is the proposed CNOT \cite{Wang_2013}
under optimization. This CNOT is based on spin of electron in a QD
trapped in a double sided optical microcavity which behaves like a
BS \cite{Hu_2009}. Inside the grey background, we consider two input
photons 1 and 2, being the control and target photons, with initial
states$\left|\Psi_{ph}^{1}\right\rangle $ and $\left|\Psi_{ph}^{2}\right\rangle $,
respectively. 

\begin{center}
\begin{figure}[H]
\begin{centering}
\includegraphics[width=0.55\columnwidth]{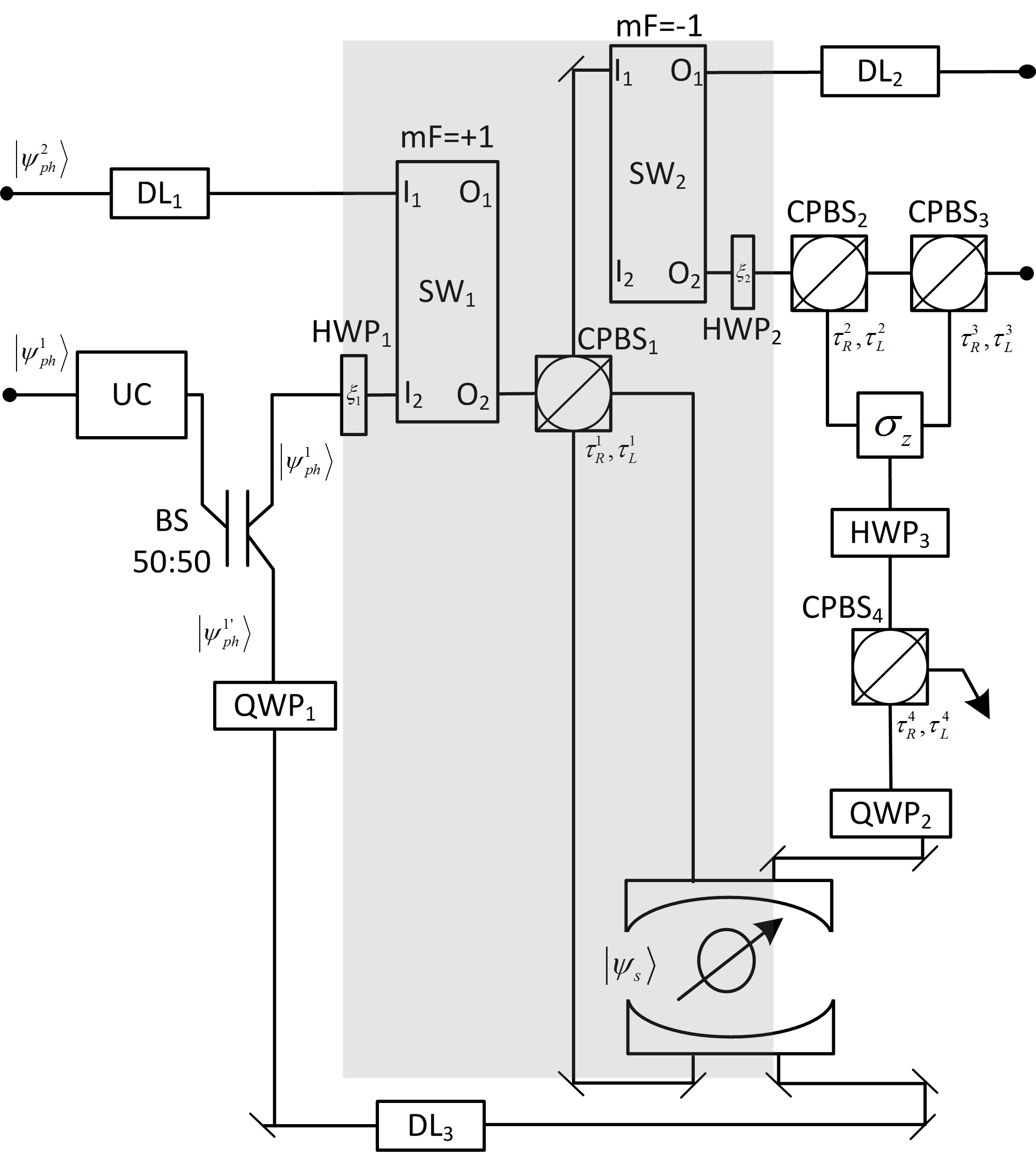}
\par\end{centering}

\caption{\label{fig:2} CNOT gate optimized model. }
\end{figure}

\par\end{center}

The electron spin state inside the QD is denoted $\left|\Psi_{s}\right\rangle $.
Consider the following initial states:

{\small{}
\begin{equation}
\begin{array}{c}
\left|\Psi_{ph}^{1}\right\rangle =\alpha\left|R_{1}\right\rangle +\beta\left|L_{1}\right\rangle \\
\Psi_{ph}^{2}=\delta\left|R_{2}\right\rangle +\gamma\left|L_{2}\right\rangle \\
\left|\Psi_{s}\right\rangle =\left(\left|\uparrow_{s}\right\rangle -\left|\downarrow_{s}\right\rangle \right)/\sqrt{2}
\end{array}\label{eq:2}
\end{equation}
}{\small \par}

The two photons come successively to interact with the optical micro-cavity.
For the coupled cavity, when considering equal the frequencies of
the input photon, cavity mode and the spin-dependent optical transition,
the reflection and transmission coefficients of the double sided optical
micro-cavity system used in the CNOT design, are denoted $r\left(\omega\right)$
and $t\left(\omega\right)$, respectively, and are given by \cite{Wang_2013,Bonato_2010,Hu_2009}:

{\small{}
\begin{equation}
t\left(\omega\right)=-\frac{2\gamma\kappa}{\gamma\left(2\kappa+\kappa_{s}\right)+4g^{2}};\,\,\,\,\,r\left(\omega\right)=1+t\left(\omega\right)\label{eq:3}
\end{equation}
}{\small \par}

where $g$ is the coupling strength, $\kappa$ and $\kappa_{s}/2$
are the cavity field decay rate into the input/output modes and the
leaky modes, respectively, and $\gamma/2$ is the $X^{-}$ dipole
decay rate. For the uncoupled cavity, the reflection and transmission
coefficients are denoted $r_{0}\left(\omega\right)$ and $t_{0}\left(\omega\right)$
, and they are directly obtained from equation \ref{eq:3} for $g=0$.
For a realistic spin cavity unit, the side leakage and cavity loss
can not be neglected. In this case, $t\left(\omega\right)$ in the
coupled cavity and $r_{0}\left(\omega\right)$ in the uncoupled cavity
will introduce bit-flip errors. Relevant energy levels and optical
selection rules for exciton $X^{-}$ inside the single charged GaAs/
InAs QD, have been well detailed in \cite{Wang_2013,Wei_2014}, and
the dynamics of the interaction of the QD spin in a double sided optical
micro-cavity, for $r_{0}=\left|r_{0}\left(\omega\right)\right|$,
$t_{0}=\left|t_{0}\left(\omega\right)\right|$, $r_{1}=\left|r\left(\omega\right)\right|$
and $t_{1}=\left|t\left(\omega\right)\right|$, are given as follows:

{\small{}
\begin{equation}
\begin{array}{c}
\left|R^{\downarrow},\,\uparrow_{s}\right\rangle \rightarrow-t_{0}\left|R^{\downarrow},\,\uparrow_{s}\right\rangle -r_{0}\left|L^{\uparrow},\,\uparrow_{s}\right\rangle \\
\left|R^{\downarrow},\,\downarrow_{s}\right\rangle \rightarrow r_{1}\left|L^{\uparrow},\,\downarrow_{s}\right\rangle +t_{1}\left|R^{\downarrow},\,\downarrow_{s}\right\rangle \\
\left|R^{\uparrow},\,\uparrow_{s}\right\rangle \rightarrow r_{1}\left|L^{\downarrow},\,\uparrow_{s}\right\rangle +t_{1}\left|R^{\uparrow},\,\uparrow_{s}\right\rangle \\
\left|R^{\uparrow},\,\downarrow_{s}\right\rangle \rightarrow-t_{0}\left|R^{\uparrow},\,\downarrow_{s}\right\rangle -r_{0}\left|L^{\downarrow},\,\downarrow_{s}\right\rangle \\
\left|L^{\downarrow},\,\uparrow_{s}\right\rangle \rightarrow r_{1}\left|R^{\uparrow},\,\uparrow_{s}\right\rangle +t_{1}\left|L^{\downarrow},\,\uparrow_{s}\right\rangle \\
\left|L^{\downarrow},\,\downarrow_{s}\right\rangle \rightarrow-t_{0}\left|L^{\downarrow},\,\downarrow_{s}\right\rangle -r_{0}\left|R^{\uparrow},\,\downarrow_{s}\right\rangle \\
\left|L^{\uparrow},\,\uparrow_{s}\right\rangle \rightarrow-t_{0}\left|L^{\uparrow},\,\uparrow_{s}\right\rangle -r_{0}\left|R^{\downarrow},\,\uparrow_{s}\right\rangle \\
\left|L^{\uparrow},\,\downarrow_{s}\right\rangle \rightarrow r_{1}\left|R^{\downarrow},\,\downarrow_{s}\right\rangle +t_{1}\left|L^{\uparrow},\,\downarrow_{s}\right\rangle 
\end{array}\label{eq:45}
\end{equation}
}{\small \par}

Photon 1 first passes through HWP1, then it travels through the optical
micro-cavity and then it passes through HWP2. The two switches used
in the CNOT are denoted SW1 and SW2, the transmittance and reflectance
coefficients of SW1 are denoted $T_{1,2}^{1}$, $T_{2,1}^{1}$, $R_{1,1}^{1}$
and $R_{2,2}^{1}$, while they are $T_{1,2}^{2}$, $T_{2,1}^{2}$,
$R_{1,1}^{2}$ and $R_{2,2}^{2}$ for SW2. No details were given about
SW1 and SW2 in \cite{Wang_2013}, and they were supposed to switch
between photons 1 and 2 perfectly. After a certain time defined by
DL1, photon 2 is switched by SW1 and injected to the spin cavity system,
but before entering and after leaving the system, two Hadamard transforms
are performed on the electron spin state, through $\nicefrac{\pi}{2}$
microwave pulses \cite{Wang_2013,Bonato_2010}, which transforms the
state {\footnotesize{}$\left|\uparrow_{s}\right\rangle \rightarrow\left(\left|\uparrow_{s}\right\rangle +\left|\downarrow_{s}\right\rangle \right)/\sqrt{2}$}
and{\footnotesize{} $\left|\downarrow_{s}\right\rangle \rightarrow\left(\left|\uparrow_{s}\right\rangle -\left|\downarrow_{s}\right\rangle \right)/\sqrt{2}$}.
After being switched by SW2, photon 2 is delayed by DL2 to wait for
the interaction between photon 1 and it's clone.

For the inputs of equation \ref{eq:2}, the state at the output of
the CNOT is then transformed as follows:

{\small{}
\begin{equation}
\begin{array}{c}
\left|\Psi_{ph}^{1}\right\rangle \otimes\left|\Psi_{ph}^{2}\right\rangle \otimes\left|\Psi_{s}\right\rangle \rightarrow\\
\left(\alpha\delta\left|R_{1}\right\rangle \left|R_{2}\right\rangle +\alpha\gamma\left|R_{1}\right\rangle \left|L_{2}\right\rangle -\beta\delta\left|L_{1}\right\rangle \left|L_{2}\right\rangle -\beta\gamma\left|L_{1}\right\rangle \left|R_{2}\right\rangle \right)\left|\uparrow_{s}\right\rangle \\
+\left(\alpha\delta\left|R_{1}\right\rangle \left|R\right\rangle _{2}+\alpha\gamma\left|R_{1}\right\rangle \left|L_{2}\right\rangle +\beta\delta\left|L_{1}\right\rangle \left|L_{2}\right\rangle +\beta\gamma\left|L_{1}\right\rangle \left|R_{2}\right\rangle \right)\left|\downarrow_{s}\right\rangle 
\end{array}\label{eq:4}
\end{equation}
}{\small \par}

It is clear from equation \ref{eq:4} that the CNOT function is correctly
entangled to the spin state $\left|\downarrow_{s}\right\rangle $,
but a $\left(-\right)$ sign is introduced to the CNOT when photon
1 is in the state $\left|L_{1}\right\rangle $ and both photons are
entangled with the spin state $\left|\uparrow_{s}\right\rangle $.
A measurement of the spin is required to determine the spin state,
and then decide whether to apply an identity $\left(I\right)$ or
a negation gate ($\sigma_{z}$) on photon 1, to get correct CNOT entangled
with both $\left|\uparrow_{s}\right\rangle $ and $\left|\downarrow_{s}\right\rangle $
states. This heralded function has a fidelity of 94 \%, which means
that the correct CNOT performed only with 47 \% of success probability.
However, without spin measurement, this CNOT model cannot be used
in a serial or parallel combination, since the success probability
of the entire circuit will decrease exponentially. 

Our main idea is to eliminate this $\left(-\right)$ sign at the output,
in order to be independent of the spin state for further circuit realization.
The idea is to apply a $\sigma_{z}$ transform on photon 1, being
at the state $\left|L_{1}\right\rangle $, only when the spin state
at the output is $\left|\uparrow_{s}\right\rangle $. A measurement
of the spin state inside a QD has been addressed in \cite{Hu_2009}:
if we have an horizontally-polarized $\left(\left|H\right\rangle \right)$
or vertically-polarized $\left(\left|V\right\rangle \right)$ single
photon at the input of QD spin system, being initially at the state
$\mu\left|\uparrow_{s}\right\rangle +\nu\left|\downarrow_{s}\right\rangle $,
it is possible using a QWP after the QD system to transmit the state
of the electron to the photon as $\mu\left|R\right\rangle +\nu\left|L\right\rangle $.
Based on this idea, we use in our architecture a UC to clone photon
1 and produce another photon 1' in the same state $\left|\Psi_{ph}^{1'}\right\rangle =\left|\Psi_{ph}^{1}\right\rangle =\alpha\left|R_{1'}\right\rangle +\beta\left|L_{1'}\right\rangle $.
\foreignlanguage{canadian}{At the output of the UC, photon 1 and photon
1' are separated by a 50;50 BS. Photon 1 remains the control photon
of the CNOT and photon 1' is created to be further used as control
for the }$\sigma_{z}$ gate. After traversing QWP1, the state of photon
1' becomes $\left|\Psi_{ph}^{1'}\right\rangle =\alpha\left|H_{1'}\right\rangle +\beta\left|V_{1'}\right\rangle $.
Photon 1' is then delayed by DL3 to wait for photons 1 and 2 to pass
the QD system and alter the spin state (the spin state is initially
given by equation \ref{eq:2}, and after two imperfect Hadamard gates,
it becomes $\mu\left|\uparrow_{s}\right\rangle +\nu\left|\downarrow_{s}\right\rangle $,
for $\mu\approx\nu\approx\nicefrac{1}{2}$). Photon 1' passes through
the QD spin system and after QWP2, it's state becomes $\left|\Psi_{ph}^{1'}\right\rangle =\mu\left|R_{1}\right\rangle +\nu\left|L_{1}\right\rangle $.
CPBS4 transmits $\mu\left|R_{1}\right\rangle $ while $\nu\left|L_{1}\right\rangle $
is discarded. The transmitted $\mu\left|R_{1}\right\rangle $ is flipped
to $\mu\left|L_{1}\right\rangle $ by HWP3. At this level, photon
1' is present with same probability amplitude $\mu$ of the electron
spin being at the state $\left|\uparrow_{s}\right\rangle $, moreover,
it is exactly in the same mode of photon 1, this allows it to serve
as control for the $\sigma_{z}$ gate. This gate should perform a
$\left(-\right)$ sign only to photon 1 being at the state $\left|L_{1}\right\rangle $,
this is the role of CPBS2 and CPBS3. Finally, the time interval between
all photons, the paths lengths traveled by photons and the time delay
of DL1, DL2 and DL3, should take into consideration cavity photon
lifetime and single charged electron spin coherence time \cite{Wang_2013}. 

We consider $\xi_{1}$ and $\xi_{2}$ to be the errors related to
HWP1 and HWP2. We suppose that $\left(\tau_{R}^{1},\,\tau_{L}^{1}\right)$,$\left(\tau_{R}^{2},\,\tau_{L}^{2}\right)$,
$\left(\tau_{R}^{3},\,\tau_{L}^{3}\right)$ and $\left(\tau_{R}^{4},\,\tau_{L}^{4}\right)$
are the errors related to CPBS1, CPBS2, CPBS3 and CPBS4, respectively.
For simplicity, we neglect errors due to QWP1, QWP2 and HWP3. For
the same inputs of equation \ref{eq:2}, we compute the output of
the optimized CNOT and we obtain:

\begin{singlespace}
{\small{}
\begin{equation}
\begin{array}{c}
\left|\Psi_{ph}^{1}\right\rangle \otimes\left|\Psi_{ph}^{2}\right\rangle \otimes\left|\Psi_{s}\right\rangle \rightarrow\sqrt{T_{1,2}^{1}\,R_{2,2}^{1}\,T_{1,2}^{2}\,R_{1,1}^{2}\,F_{cloner}}\\
\times\left(\left(\eta_{1}\left|R_{1}\right\rangle \left|R_{2}\right\rangle +\eta_{2}\left|R_{1}\right\rangle \left|L_{2}\right\rangle +\eta_{3}\left|L_{1}\right\rangle \left|L_{2}\right\rangle +\eta_{4}\left|L_{1}\right\rangle \left|R_{2}\right\rangle \right)\left|\uparrow_{s}\right\rangle \right.\\
\left.+\left(\eta_{5}\left|R_{1}\right\rangle \left|R_{2}\right\rangle +\eta_{6}\left|R_{1}\right\rangle \left|L_{2}\right\rangle +\eta_{7}\left|L_{1}\right\rangle \left|L_{2}\right\rangle +\eta_{8}\left|L_{1}\right\rangle \left|R_{2}\right\rangle \right)\left|\downarrow_{s}\right\rangle \right)
\end{array}\label{eq:5}
\end{equation}
}{\small \par}

{\small{}where:}{\small \par}
\end{singlespace}

{\small{}
\begin{equation}
\begin{array}{c}
\eta_{1}=\frac{\sqrt{1-\xi_{2}}}{2\sqrt{2}}\left(\left(a_{2}a_{2}^{'}+a_{4}a_{4}^{'}-a_{1}a_{1}^{'}-a_{3}a_{3}^{'}\right)\left(\delta\delta^{'}+\gamma\gamma^{'}\right)+\right.\\
\left.\left(a_{2}a_{2}^{"}+a_{4}a_{4}^{"}-a_{1}a_{1}^{"}-a_{3}a_{3}^{"}\right)\left(\delta\delta^{"}+\gamma\gamma^{"}\right)\right)\\
a_{1}=\left(\alpha+\beta\right)\sqrt{\left(1-\tau_{R}^{1}\right)\left(1-\xi_{1}\right)/2}\\
a_{2}=\left(\alpha+\beta\right)\sqrt{\tau_{R}^{1}\left(1-\xi_{1}\right)/2}\\
a_{3}=\left(\alpha-\beta\right)\sqrt{\left(1-\tau_{L}^{1}\right)\left(1+\xi_{1}\right)/2}\\
a_{4}=\left(\alpha-\beta\right)\sqrt{\tau_{L}^{1}\left(1+\xi_{1}\right)/2}\\
a_{1}^{'}=\sqrt{\left(1-\tau_{R}^{1}\right)}\left(t_{0}+t_{1}\right)+\sqrt{\left(1-\tau_{L}^{1}\right)}\left(r_{0}+r_{1}\right)\\
a_{2}^{'}=\sqrt{\tau_{R}^{1}}\left(t_{0}+t_{1}\right)+\tau_{L}^{1}\left(r_{0}+r_{1}\right)\\
a_{3}^{'}=\sqrt{\left(1-\tau_{R}^{1}\right)}\left(r_{0}+r_{1}\right)+\sqrt{\left(1-\tau_{L}^{1}\right)}\left(t_{0}+t_{1}\right)\\
a_{4}^{'}=\sqrt{\tau_{R}^{1}}\left(r_{0}+r_{1}\right)+\sqrt{\tau_{L}^{1}}\left(t_{0}+t_{1}\right)\\
a_{1}^{"}=\sqrt{\left(1-\tau_{R}^{1}\right)}\left(t_{0}-t_{1}\right)+\sqrt{\left(1-\tau_{L}^{1}\right)}\left(r_{0}-r_{1}\right)\\
a_{2}^{"}=\sqrt{\tau_{R}^{1}}\left(t_{1}-t_{0}\right)+\tau_{L}^{1}\left(r_{1}+r_{0}\right)\\
a_{3}^{"}=\sqrt{\left(1-\tau_{R}^{1}\right)}\left(r_{0}-r_{1}\right)+\sqrt{\left(1-\tau_{L}^{1}\right)}\left(t_{0}-t_{1}\right)\\
a_{4}^{"}=\sqrt{\tau_{R}^{1}}\left(r_{1}+r_{0}\right)+\sqrt{\tau_{L}^{1}}\left(t_{1}-t_{0}\right)\\
\delta^{'}=t_{1}\tau_{R}^{1}-t_{0}\left(1-\tau_{R}^{1}\right)\\
\gamma^{'}=r_{1}\sqrt{\tau_{R}^{1}\tau_{L}^{1}}-r_{0}\sqrt{\left(1-\tau_{R}^{1}\right)\left(1-\tau_{L}^{1}\right)}\\
\delta^{"}=t_{1}\left(1-\tau_{R}^{1}\right)-t_{0}\tau_{R}^{1}\\
\gamma^{"}=r_{1}\sqrt{\left(1-\tau_{R}^{1}\right)\left(1-\tau_{L}^{1}\right)}-r_{0}\sqrt{\tau_{R}^{1}\tau_{L}^{1}}
\end{array}\label{eq:6}
\end{equation}
}{\small \par}

and $\left\{ \eta_{i}\right\} _{2\leq i\leq8}$ have all the same
form of equation \ref{eq:6}. 

It is worth highlighting the fact that {\small{}$\Theta=\left(-1\right)\times\sqrt{\left(1-\tau_{L}^{2}\right)\left(1-\tau_{L}^{3}\right)\left(1-\tau_{R}^{4}\right)}$
i}s the operator that will eliminate the (-) sign for the CNOT being
entangled with $\left|\uparrow_{s}\right\rangle $, this operator
appears only in $\eta_{3}$ and $\eta_{4}$ of equation \ref{eq:5}.
This is the main contribution in this work since the CNOT function
is correctly entangled with both $\left|\uparrow_{s}\right\rangle $
and $\left|\downarrow_{s}\right\rangle $ spin states. To measure
the performance of our optimized CNOT, we refer to the fidelity denoted
$F_{CNOT}$ and given by \cite{Wang_2013}:

{\small{}
\begin{equation}
F_{CNOT}=\left\langle \overline{\Psi_{in}|U_{CNOT}^{\dagger}\rho_{t}U_{CNOT}|\Psi_{in}}\right\rangle \label{eq:9}
\end{equation}
}{\small \par}

where the upper line indicates that the fidelity is obtained according
to the average over all possible four input states $\left|\Psi_{in}\right\rangle $,
$U_{CNOT}$ is the ideal CNOT transform, $\rho_{t}=\left|\Psi_{out}\right\rangle \left\langle \Psi_{out}\right|$,
with $\left|\Psi_{out}\right\rangle $ is the state at the output
of the CNOT for the specific $\left|\Psi_{in}\right\rangle $ input.

\section{Simulation results and experimental challenges}

A first simulation concerns only the original proposed CNOT, where
we study the impact of the errors of HWP1, HWP2 and CPBS1. To this
end, we consider perfect SW1 and SW2{\small{} $\left(T_{1,2}^{1}=T_{1,2}^{2}=R_{2,2}^{1}=R_{1,1}^{2}=1\right)$},
and we vary all errors around a realistic range of $10^{-2}$.

\begin{figure}[H]
\begin{centering}
\subfloat[\label{fig:3a}]{\begin{centering}
\includegraphics[width=0.35\columnwidth]{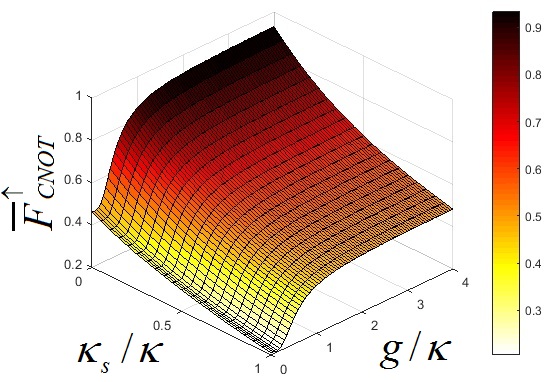}
\par\end{centering}

}\subfloat[\label{fig:3b}]{\begin{centering}
\includegraphics[width=0.35\columnwidth]{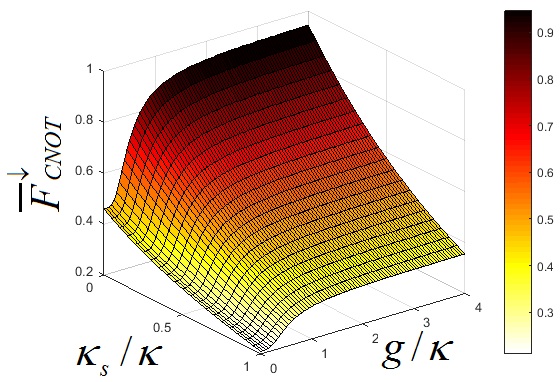}
\par\end{centering}

}
\par\end{centering}

\caption{\label{fig:3}The average fidelity of the photonic CNOT gate versus
the normalized coupling strengths $\kappa_{s}/\kappa$ and $g/\kappa$.
(a) The fidelity for electron spin in the state $\left|\uparrow_{s}\right\rangle $.
(b) The fidelity for electron spin in the state $\left|\downarrow_{s}\right\rangle $.}
\end{figure}

We illustrate in figures \ref{fig:3a} and \ref{fig:3b}, the average
fidelities for spin $\left|\uparrow_{s}\right\rangle $ and $\left|\downarrow_{s}\right\rangle $
states, denoted $\overline{F}_{CNOT}^{\uparrow}$ and $\overline{F}_{CNOT}^{\downarrow}$,
versus the normalized coupling strength. Here we have set $\gamma=0.1\kappa$. 

In \cite{Wang_2013}, errors due to HWP1, HWP2 and CPBS1 have been
neglected and it has been shown that the CNOT provides best $\overline{F}_{CNOT}^{\uparrow}$
or $\overline{F}_{CNOT}^{\downarrow}$ value around 93.74\% for the
strong coupling regime (obtained for $g>\left(\kappa_{s}+\kappa\right)/4$),
and 32.34 \% for the weak coupling regime (obtained when $g<\left(\kappa_{s}+\kappa\right)/4$).
If we consider the same parameters used for the strong coupling ($\kappa_{s}=0.05\kappa$
and $g=2.5\kappa$) and weak coupling ($\kappa_{s}=1.0\kappa$ and
$g=0.45\kappa$), our simulation shows that the errors affect the
fidelities $\overline{F}_{CNOT}^{\uparrow}$ and $\overline{F}_{CNOT}^{\downarrow}$
, and we obtain best values around 87.89\% and 30.02\%, respectively.

\begin{figure}[H]
\begin{centering}
\subfloat[\label{fig:4a}]{\begin{centering}
\includegraphics[width=0.35\columnwidth]{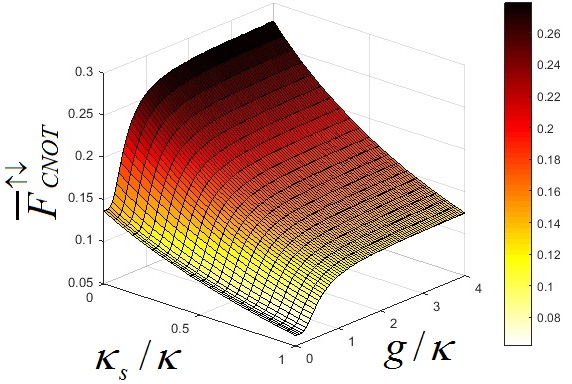}
\par\end{centering}

}\subfloat[\label{fig:4b}]{\begin{centering}
\includegraphics[width=0.35\columnwidth]{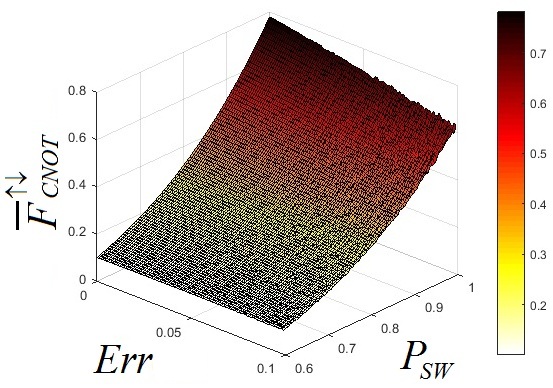}
\par\end{centering}

}
\par\end{centering}

\caption{\label{fig:4}Simulation of the fidelity of the CNOT with realistic
errors. (a) The average fidelity versus the normalized coupling strengths
$\kappa_{s}/\kappa$ and $g/\kappa$. (b) The average fidelity versus
errors and success probability of the SWs.}
\end{figure}

Another simulation concerns our optimized model while taking into
consideration realistic features of all devices. In this case, we
consider SW1 and SW2 realized according to \cite{Shomroni_2014}.
For SW1 being initially in the state $m_{F}=+1$, the experimental
results obtained are $T_{1,2}^{1}=89.9\%$, $R_{2,2}^{1}=65\%$. For
SW2 being initially in the ground state $m_{F}=-1$, the obtained
coefficients are $T_{1,2}^{2}=95.6\%$ and $R_{1,1}^{2}=64.8\%$.
We consider the universal cloning of polarization state experimentally
realized and providing $F_{cloner}=0.82$ \cite{Fasel_2002}. We consider
arbitrary errors around $10^{-2}$ affecting separately all devices
of the CNOT of figure \ref{fig:2} (except QWP1, QWP2 and HWP3). With
these assumptions, we show in Figure \ref{fig:4a} the average fidelity
of the CNOT function being correctly entangled with both $\left|\uparrow_{s}\right\rangle $
and $\left|\downarrow_{s}\right\rangle $ , denoted $\overline{F}_{CNOT}^{\uparrow\downarrow}$.
Best values according to figure \ref{fig:4a} in the strong coupling
regime is $\overline{F}_{CNOT}^{\uparrow\downarrow}=26.27\%$.

It is clear that $\overline{F}_{CNOT}^{\uparrow\downarrow}$ value
is highly sensitive to the fidelity of the cloner and SW1 and SW2
imperfections, we have considered only the strong coupling regime
and the optimal cloner with $F_{UC}=5/6$. We denote $E_{rr}$ the
set of all errors affecting the CNOTs components and we vary them
in $\left[10^{-4}..10^{-1}\right]$. We consider also the same range
of errors separately altering the coefficients $T_{1,2}^{1}$, $T_{1,2}^{2}$,
$R_{2,2}^{1}$ and $R_{1,1}^{2}$, therefore, we denote $P_{SW}\simeq T_{1,2}^{1}$$\simeq T_{1,2}^{2}$$\simeq R_{2,2}^{1}$$\simeq R_{1,1}^{2}$,
and we illustrate $\overline{F}_{CNOT}^{\uparrow\downarrow}$ depending
on $E_{rr}$ and $P_{SW}$ in figure \ref{fig:4b}. The best fidelity
permitted by our CNOT for lowest error range and $P_{SW}$ approaching
unity is $\overline{F}_{CNOT}^{\uparrow\downarrow}$=78\%. This fidelity
is very close to $F_{UC}$ and our optimized CNOT is very advantageous
since neither a measurement of the electron spin state nor an extra
treatment are required to allow using the CNOT outputs as inputs for
another circuits.

\section{Conclusion}

We propose a quantum CNOT gate that overcomes the inefficiencies of
a previously published CNOT design based on Quantum-Dot system. The
previous proposal provides fidelity of 94 \% but with a heralded functioning,
which means that the correct CNOT performed only with 47 \% probability
success. Our CNOT functioning is not heralded by any spin states and
it provides a success of 78\%. This is a highly innovative result
since it uses the quantum Cloner. This design will lead to another
CNOT that uses the cloner with better fidelity approaching the cloner
optimal limit of 5/6 and will allow possible generalization of the
CNOT to all $C^{n}NOT$ photonic gates. 

\bibliographystyle{spiebib}
\addcontentsline{toc}{section}{\refname}\bibliography{Quantum_Bib_2019}

\begin{thebibliography}{10}

\bibitem{Kok_2007}
P.~Kok, W.~J. Munro, K.~Nemoto, T.~C. Ralph, J.~P. Dowling, and G.~J. Milburn,
  ``Linear optical quantum computing with photonic qubits,'' {\em Rev. Mod.
  Phys.}~{\bf 79}, pp.~135--174, Jan 2007.

\bibitem{Shende_2006}
V.~V. Shende, S.~S. Bullock, and I.~L. Markov, ``Synthesis of quantum-logic
  circuits,'' {\em IEEE Transactions on Computer-Aided Design of Integrated
  Circuits and Systems}~{\bf 25}, pp.~1000--1010, June 2006.

\bibitem{Maslov_2008}
D.~Maslov, G.~W. Dueck, D.~M. Miller, and C.~Negrevergne, ``Quantum circuit
  simplification and level compaction,'' {\em IEEE Transactions on
  Computer-Aided Design of Integrated Circuits and Systems}~{\bf 27},
  pp.~436--444, March 2008.

\bibitem{Pittman_2001}
T.~B. Pittman, B.~C. Jacobs, and J.~D. Franson, ``Probabilistic quantum logic
  operations using polarizing beam splitters,'' {\em Phys. Rev. A}~{\bf 64},
  p.~062311, Nov 2001.

\bibitem{Ralph_2002}
T.~C. Ralph, N.~K. Langford, T.~B. Bell, and A.~G. White, ``Linear optical
  controlled-not gate in the coincidence basis,'' {\em Phys. Rev. A}~{\bf 65},
  p.~062324, Jun 2002.

\bibitem{OBrien_2003}
J.~L. O'Brien, G.~J. Pryde, A.~G. White, T.~C. Ralph, and D.~Branning,
  ``Demonstration of an all-optical quantum controlled-not gate,'' {\em
  Nature}~{\bf 426}, pp.~264 EP --, Nov 2003.

\bibitem{Pittman_2003}
T.~B. Pittman, M.~J. Fitch, B.~C. Jacobs, and J.~D. Franson, ``Experimental
  controlled-not logic gate for single photons in the coincidence basis,'' {\em
  Phys. Rev. A}~{\bf 68}, p.~032316, Sep 2003.

\bibitem{Gasparoni_2004}
S.~Gasparoni, J.-W. Pan, P.~Walther, T.~Rudolph, and A.~Zeilinger,
  ``Realization of a photonic controlled-not gate sufficient for quantum
  computation,'' {\em Phys. Rev. Lett.}~{\bf 93}, p.~020504, Jul 2004.

\bibitem{Pittman_2004}
T.~B. Pittman, B.~C. Jacobs, and J.~D. Franson, ``Probabilistic quantum encoder
  for single-photon qubits,'' {\em Phys. Rev. A}~{\bf 69}, p.~042306, Apr 2004.

\bibitem{Bao_2007}
X.-H. Bao, T.-Y. Chen, Q.~Zhang, J.~Yang, H.~Zhang, T.~Yang, and J.-W. Pan,
  ``Optical nondestructive controlled-not gate without using entangled
  photons,'' {\em Phys. Rev. Lett.}~{\bf 98}, p.~170502, Apr 2007.

\bibitem{Clark_2009}
A.~S. Clark, J.~Fulconis, J.~G. Rarity, W.~J. Wadsworth, and J.~L. O'Brien,
  ``All-optical-fiber polarization-based quantum logic gate,'' {\em Phys. Rev.
  A}~{\bf 79}, p.~030303, Mar 2009.

\bibitem{Gueddana_2013a}
A.~Gueddana, M.~Attia, and R.~Chatta, ``Non-deterministic quantum cnot gate
  with double encoding,'' {\em Proc.SPIE}~{\bf 8875}, pp.~8875 -- 8875 -- 6,
  2013.

\bibitem{Gueddana_2015}
A.~Gueddana, A.~Moez, and R.~Chatta, ``Abstract probabilistic cnot gate model
  based on double encoding: study of the errors and physical realizability,''
  {\em Proc.SPIE}~{\bf 9377}, pp.~9377 -- 9377 -- 8, 2015.

\bibitem{Knill_2003}
E.~Knill, ``Bounds on the probability of success of postselected nonlinear sign
  shifts implemented with linear optics,'' {\em Phys. Rev. A}~{\bf 68},
  p.~064303, Dec 2003.

\bibitem{Plantenberg_2007}
J.~H. Plantenberg, P.~C. de~Groot, C.~J. P.~M. Harmans, and J.~E. Mooij,
  ``Demonstration of controlled-not quantum gates on a pair of superconducting
  quantum bits,'' {\em Nature}~{\bf 447}, pp.~836 EP --, Jun 2007.

\bibitem{Luo_2016}
M.-X. Luo, H.-R. Li, and H.~Lai, ``Quantum computation based on photonic
  systems with two degrees of freedom assisted by the weak cross-kerr
  nonlinearity,'' {\em Scientific Reports}~{\bf 6}, pp.~29939 EP --, Jul 2016.
\newblock Article.

\bibitem{Li_2013}
C.-Y. Li, Z.-R. Zhang, S.-H. Sun, M.-S. Jiang, and L.-M. Liang, ``Logic-qubit
  controlled-not gate of decoherence-free subspace with nonlinear quantum
  optics,'' {\em J. Opt. Soc. Am. B}~{\bf 30}, pp.~1872--1877, Jul 2013.

\bibitem{Wei_2013}
H.-R. Wei and F.-G. Deng, ``Scalable photonic quantum computing assisted by
  quantum-dot spin in double-sided optical microcavity,'' {\em Opt.
  Express}~{\bf 21}, pp.~17671--17685, Jul 2013.

\bibitem{Wei_2014}
H.-R. Wei and F.-G. Deng, ``Universal quantum gates on electron-spin qubits
  with quantum dots inside single-side optical microcavities,'' {\em Opt.
  Express}~{\bf 22}, pp.~593--607, Jan 2014.

\bibitem{Wang_2013}
H.-F. Wang, J.-J. Wen, A.-D. Zhu, S.~Zhang, and K.-H. Yeon, ``Deterministic
  cnot gate and entanglement swapping for photonic qubits using a quantum-dot
  spin in a double-sided optical microcavity,'' {\em Physics Letters A}~{\bf
  377}(40), pp.~2870 -- 2876, 2013.

\bibitem{Luo_2014}
M.-X. Luo and X.~Wang, ``Parallel photonic quantum computation assisted by
  quantum dots in one-side optical microcavities,'' {\em Scientific
  Reports}~{\bf 4}, pp.~5732 EP --, Jul 2014.
\newblock Article.

\bibitem{Bonato_2010}
C.~Bonato, F.~Haupt, S.~S.~R. Oemrawsingh, J.~Gudat, D.~Ding, M.~P. van Exter,
  and D.~Bouwmeester, ``Cnot and bell-state analysis in the weak-coupling
  cavity qed regime,'' {\em Phys. Rev. Lett.}~{\bf 104}, p.~160503, Apr 2010.

\bibitem{Gueddana_2018d}
A.~Gueddana and V.~Lakshminarayanan, ``Comment on 'deterministic cnot gate and
  entanglement swapping for photonic qubits using a quantum-dot spin in a
  double-sided optical microcavity' by h.f. wang et al. [physics letters a 377
  (2013) 2870-2876],'' {\em Physics Letters A} , 2018.

\bibitem{Fasel_2002}
S.~Fasel, N.~Gisin, G.~Ribordy, V.~Scarani, and H.~Zbinden, ``Quantum cloning
  with an optical fiber amplifier,'' {\em Phys. Rev. Lett.}~{\bf 89},
  p.~107901, Aug 2002.

\bibitem{Lamas_Linares_2002}
A.~Lamas-Linares, C.~Simon, J.~C. Howell, and D.~Bouwmeester, ``Experimental
  quantum cloning of single photons,'' {\em Science}~{\bf 296}(5568),
  pp.~712--714, 2002.

\bibitem{Martini_2004}
F.~D. Martini, D.~Pelliccia, and F.~Sciarrino, ``Contextual, optimal, and
  universal realization of the quantum cloning machine and of the not gate,''
  {\em Phys. Rev. Lett.}~{\bf 92}, p.~067901, Feb 2004.

\bibitem{Bartifmmodemathringuelserufiifmmodecheckselsevsfikova_2007}
L.~Bart\ifmmode \mathring{u}\else \r{u}\fi{}\ifmmode~\check{s}\else
  \v{s}\fi{}kov\'a, M.~Du\ifmmode~\check{s}\else \v{s}\fi{}ek,
  A.~\ifmmode~\check{C}\else \v{C}\fi{}ernoch, J.~Soubusta, and
  J.~Fiur\'a\ifmmode~\check{s}\else \v{s}\fi{}ek, ``Fiber-optics implementation
  of an asymmetric phase-covariant quantum cloner,'' {\em Phys. Rev.
  Lett.}~{\bf 99}, p.~120505, Sep 2007.

\bibitem{OShea_2013}
D.~O'Shea, C.~Junge, J.~Volz, and A.~Rauschenbeutel, ``Fiber-optical switch
  controlled by a single atom,'' {\em Phys. Rev. Lett.}~{\bf 111}, p.~193601,
  Nov 2013.

\bibitem{Smart_2014}
A.~G. Smart, ``A quantum switch routes photons one by one,'' {\em Physics
  Today}~{\bf 67}, pp.~9--15, 2014.

\bibitem{Sun_2016}
S.~Sun, H.~Kim, G.~S. Solomon, and E.~Waks, ``A quantum phase switch between a
  single solid-state spin and a photon,'' {\em Nature Nanotechnology}~{\bf 11},
  pp.~539 EP --, Feb 2016.
\newblock Article.

\bibitem{Volz_2012}
T.~Volz, A.~Reinhard, M.~Winger, A.~Badolato, K.~J. Hennessy, E.~L. Hu, and
  A.~Imamoglu, ``Ultrafast all-optical switching by single photons,'' {\em
  Nature Photonics}~{\bf 6}, pp.~605 EP --, Aug 2012.

\bibitem{Shomroni_2014}
I.~Shomroni, S.~Rosenblum, Y.~Lovsky, O.~Bechler, G.~Guendelman, and B.~Dayan,
  ``All-optical routing of single photons by a one-atom switch controlled by a
  single photon,'' {\em Science}~{\bf 345}(6199), pp.~903--906, 2014.

\bibitem{Hu_2009}
C.~Y. Hu, W.~J. Munro, J.~L. O'Brien, and J.~G. Rarity, ``Proposed entanglement
  beam splitter using a quantum-dot spin in a double-sided optical
  microcavity,'' {\em Phys. Rev. B}~{\bf 80}, p.~205326, Nov 2009.

\end{thebibliography}

\end{document}